Article

# Social Sustainability of Digital Transformation: Empirical Evidence from EU-27 Countries

Saeed Nosratabadi[1], Thabit Atobishi[2] and Szilárd Hegedűs [3,*]

[1] Doctoral School of Economic and Regional Sciences, Hungarian University of Agriculture and Life Sciences Gödöllő 2100, Hungary;
[2] Department of Management Information Systems, Amman Arab University, Amman 11953, Jordan;
[3] Faculty of Finance and Accountancy, Budapest Business School, Budapest 1149, Hungary

**Abstract:** In the EU-27 countries, the importance of social sustainability of digital transformation (SOSDIT) is heightened by the need to balance economic growth with social cohesion. By prioritizing SOSDIT, the EU can ensure that its citizens are not left behind in the digital transformation process and that technology serves the needs of all Europeans. Therefore, the current study aimed firstly to evaluate the SOSDIT of EU-27 countries and then to model its importance in reaching sustainable development goals (SDGs). The current study, using structural equation modeling, provided quantitative empirical evidence that digital transformation in Finland, the Netherlands, and Denmark are respectively most socially sustainable. It is also found that SOSDIT leads the countries to have a higher performance in reaching SDGs. Finally, the study provided evidence implying the inverse relationship between the Gini coefficient and reaching SDGs. In other words, the higher the Gini coefficient of a country, the lower its performance in reaching SDGs. The findings of this study contribute to the literature of sustainability and digitalization. It also provides empirical evidence regarding the SOSDIT level of EU-27 countries that can be a foundation for the development of policies to improve the sustainability of digital transformation. According to the findings, this study provides practical recommendations for countries to ensure that their digital transformation is sustainable and has a positive impact on society.

## 1. Introduction

Digital transformation is a process that is happening across various sectors in European countries. It encompasses the use of digital technologies, such as the internet, mobile devices, big data and analytics, and artificial intelligence, to improve the way organizations and governments operate and deliver services to citizens (Aly 2022). The European Union has made digital transformation a priority and has implemented policies and initiatives to drive digitalization across the member states. This includes the Digital Single Market strategy, which aims to remove barriers to online trade and create a level playing field for businesses across the EU. Additionally, the European Commission also introduced the European Data Strategy and the European Artificial Intelligence Strategy to promote the use of data and AI for economic growth and societal benefits. The European Commission considers the targets of "more than 90% of SMEs reach at least a basic level of digital intensity" and "75% of EU companies using cloud/AI/big data" for the transition to digitalization by 2030 (Commission 2021). In specific sectors, digital transformation has had a significant impact. In the healthcare sector, for example, digitalization has led to the development of telemedicine and e-health services, which allow patients to receive medical treatment remotely and improve access to healthcare for citizens in rural areas (Gjellebæk et al. 2020). In the manufacturing sector, Industry 4.0 technologies (Frank et al. 2019), such as the internet of things (Haghnegahdar et al. 2022) and advanced robotics (Parmar et al. 2022), are being used to increase efficiency and re-duce costs. In the field of education, digital transformation has led to the development of online learning platforms and the incorporation of digital tools in the classroom, which can improve access to education and personalize learning for students (Sousa et al. 2022).

Another target of the European Commission for digital transformation by 2030 is for 80% of the population to have at least some digital skills, using the slogan "gigabit for all and 5G everywhere" (Commission 2021), because 40% of Europeans do not have basic digital skills. This highlights the need to ensure that all citizens have access to digital technologies and the skills to use them, in order to participate in the digital economy and benefit from digitalization. Automation and digitalization are expected to lead to the displacement of up to 14% of jobs in the EU by the end of the decade. This highlights the need to address the potential negative impacts of digitalization on employment and to ensure that workers are reskilled to adapt to the changing labor market. These statistics demonstrate the need to address the social sustainability of digital transformation (SOSDIT) in Europe, to ensure that digitalization benefits all citizens, and that technology is used to improve the lives of all members of society, rather than exacerbating existing social inequalities. SOSDIT, indeed, is about considering the social impact of technology in the process of digital transformation.

Social sustainability in the context of digital transformation refers to the impact that technology and digitalization have on society as a whole. Iqbal et al. (Iqbal et al. 2021) define social sustainability as "a measure of the human's welfare". This includes ensuring that the benefits of digitalization are equitably distributed, and that technology is used to improve the lives of all members of society, rather than exacerbating existing social inequalities. Additionally, social sustainability in digital transformation also includes addressing the potential negative impacts of technology on employment and privacy. While digital transformation is seen as an opportunity to drive economic growth and improve citizens' lives, it is important to consider the SOSDIT in European countries to ensure that the benefits of digitalization are equitably distributed, and that technology is used to improve the lives of all members of society.

Although both European countries and institutions influencing the development of European countries (such as the European Union and the European Commission) have understood the importance of SOSDIT and have defined goals in their development plans towards achieving SOSDIT, there are no criteria and metrics to evaluate the level of SOSDIT of a country. Hence, the fundamental question is:

The first research question (RQ1): How socially sustainable is digital transformation across the EU-27 countries?

On the other hand, achieving the Sustainable Development Goals (SDGs) is important for European countries. The SDGs provide a framework for addressing some of the most pressing global challenges, such as poverty, inequality, and climate change (Clemente-Suárez et al. 2022). By achieving the SDGs, European countries can contribute to creating a more sustainable and equitable world (D'Adamo et al. 2022). The SDGs are relevant to the economic, social, and environmental challenges that European countries are facing. Achieving the SDGs can help to drive economic growth, improve citizens' lives, and create more inclusive and resilient societies. Therefore, the second research question (RQ2) is:

RQ2: Does SOSDIT lead EU-27 countries in achieving sustainable development goals?

The present study was, in fact, conducted to answer these two research questions (i.e., RQ1 and RQ2). For this purpose, the current research aims to bridge the gap in the literature by developing a conceptual model to provide a tool for measuring SOSDIT and on the other hand, to provide quantitative empirical evidence to answer the questions raised. Therefore, the current article pursues two objectives: (1) to provide a model for assessing the SOSDIT at a country level and (2) to evaluate the effect of countries' performance in SOSDIT on their achievement of SDGs.

The findings of this article not only theoretically contribute to the research literature of sustainability and digital transformation, but also provide quantitative empirical evidence to evaluate the SOSDIT of EU-27 countries. To do so, in the second section of this article, the subject literature as well as the development of hypotheses and the design of the conceptual model of this article have been elaborated. The third section of this article is dedicated to data collection and methodology applied for data analysis. The results of the quantitative analysis of the conceptual model as well as the test of the hypotheses are given in the fourth section of the article. The analysis, the implementation, and the limitations of the findings are presented in Sections 5 (i.e., Findings and Discussion) and 6 (Conclusion).

## 2. Theoretical Background

Technology has a profound impact on society, and it is essential to ensure that technology is developed, deployed, and used in ways that promote social well-being and support human values (Felt 2022). Failure to consider the social dimension of digital transformation can lead to negative consequences, such as digital divides (Reggi and Gil-Garcia 2021), unequal access to technology (Tiku 2021), data breaches (Seh et al. 2020), job losses (Bertani et al. 2020), cultural homogenization (Reid 2006), and erosions of democratic governance (Clarke and Dubois 2020). By addressing the social dimensions of digital transformation, society can ensure that technology is used to promote human development and support social progress. In other words, social sustainability of digital transformation should be able to evaluate the impact of technology on society and the ways in which society can ensure that the benefits of technology are accessible to everyone.

Social sustainability refers to the maintenance and promotion of the well-being and quality of life of individuals and communities, with a focus on ensuring that social benefits and opportunities are equitably distributed and maintained over time (Afshari et al. 2022). In order to explain social sustainability, the researchers state different dimensions and aspects, of which four have been the most referred to, that are: (1) social inclusion (Clube and Tennant 2022; Mirzoev et al. 2022), (2) human rights protection (Lozano 2022; Treviño-Lozano 2022), and (3) access to education (Leite 2022; Singh and Singh 2022).

Social inclusion refers to the active engagement of all individuals and groups in society, regardless of their background, identity, or circumstances (Fante et al. 2022). This includes ensuring equal access to resources, services, and opportunities, as well as promoting diversity and reducing discrimination and prejudice. In the digital age, access to technology and the internet has the potential to greatly improve social inclusion by connecting people and providing access to information and resources that were previously out of reach. At the same time, however, the digital divide and unequal access to technology can deepen existing inequalities and exclusions, so it is important to ensure that everyone has access to the benefits of digital transformation. Hence, the concept of digital inclusion has been developed.

Digital inclusion refers to the equal access and meaningful use of information and communication technologies (ICTs) by all members of society, regardless of age, gender, education, income, or other factors (Chohan and Hu 2022). The goal of digital inclusion is to ensure that everyone can participate fully in the digital economy and society, and that the benefits of digital technologies are shared equitably. This includes ensuring access to the internet, digital devices, digital literacy skills, and digital content and services that meet the diverse needs of individuals and communities. Digital inclusion also aims to address the digital divide, which refers to the unequal distribution of technology and its benefits, and to ensure that everyone has the opportunity to participate in the digital world and benefit from its opportunities (Aissaoui 2022). Therefore, it can be concluded that digital inclusion is one of the main aspects of SOSDIT. In fact, digital inclusion ensures equal access to technology and digital literacy skills for all members of society, which is important for bridging the digital divide and promoting equal opportunities. Accordingly, the first hypothesis of the current research is designed as follows:

**H1**. *Digital inclusion is one of the factors of SOSDIT*.

Human rights protection is an essential aspect of social sustainability, as it ensures that all individuals are treated with dignity, respect, and fairness, and have the freedom to participate in the decisions that affect their lives (Knebel et al. 2022). This includes the protection of civil, political, social, and economic rights, as well as the right to participate in the democratic process (Stone Sweet and Sandholtz 2023).

On the other hand, digital transformation has significant implications for privacy, freedom of speech, and other human rights, and it is important to ensure that these rights are protected in the digital space (Kirchschlaeger 2019). For example, the collection and use of personal data, the impact of algorithmic decision-making, and the influence of misinformation and propaganda all raise important human rights concerns (Bharti and Aryal 2022). Besides, digital transformation has created new forms of risk and harm, such as online harassment and abuse (Francisco and Felmlee 2022), cyberbullying (Giumetti and Kowalski 2022), and exposure to harmful content (Donaldson et al. 2022; Katsaros et al. 2022). At the same time, digital technologies also offer new opportunities for promoting safety, such as by providing access to emergency services, enabling new forms of community support and resilience, and promoting digital literacy and awareness of online risks. When a country has strong privacy, data protection, and security laws and regulations in place, it means that citizens' personal information is protected from unauthorized access and misuse (Rusakova et al. 2020). This helps to ensure that citizens feel safe and secure when using digital services and technologies and can trust that their personal information will not be misused. Strong privacy, data protection, and security laws and regulations can also help to prevent discrimination and bias, as well as protect citizens from fraud and identity theft. Digital privacy and security are crucial for protecting personal data and information from misuse and unauthorized access, which is essential for maintaining trust in technology and safeguarding fundamental human rights. Therefore, the second hypothesis of this study can be developed as follows:

**H2**. *Digital privacy and security is one of the factors of SOSDIT*.

Access to education is crucial for social sustainability, as it provides individuals with the skills, knowledge, and perspectives necessary to participate in the modern world and contribute to the betterment of their communities (Ahel and Lingenau 2020). Education also supports economic growth, social mobility, and improved health outcomes. Digital skills are the ability to use digital technologies effectively, efficiently, and responsibly to find, evaluate, use, create, and communicate information (Morte-Nadal and Esteban-Navarro 2022; Tinmaz et al. 2022). In order to participate in the digital economy and society, individuals need access to education and training that will help them develop digital skills (Ahel and Lingenau 2020). This includes not only formal education, but also training and support provided by employers, community organizations, and government agencies. Access to education provides individuals with the opportunity to develop digital skills, and digital skills are essential for accessing educational opportunities and benefiting from the digital transformation of education (Haleem et al. 2022). Therefore, the digital skills variable is a critical component of social sustainability of digital transformation. Digital skills address the impact of technology on employment and promote reskilling and upskilling to prepare workers for the digital economy, which is vital for supporting economic growth and social well-being (van Laar et al. 2019). Thus, the third hypothesis of this study is presented as follows:

**H3**. *Digital skills is one of the factors of SOSDIT*.

These aspects of social sustainability are important considerations in shaping the impact of digital transformation on society. To ensure that the benefits of digital technologies are equitably distributed and that the risks are mitigated, it is important to consider these aspects in the design and implementation of digital solutions.

Sustainable digital transformation is an important aspect of achieving the Sustainable Development Goals (SDGs) set by the United Nations. The SDGs are a universal call to action to end poverty, protect the planet and ensure that all people enjoy peace and prosperity by 2030. Digital technologies are seen as a key enabler to achieve these goals, by improving access to information, education, healthcare, and economic opportunities, as well as by improving the efficiency and effectiveness of various sectors. Digital technologies have the potential to contribute significantly to achieving several of the Sustainable Development Goals (SDGs) set by the United Nations. For instance, SDG 1 (Kelikume 2021): No Poverty can be advanced by providing access to financial services and digital skills training to underserved communities, thereby creating new economic opportunities for individuals and communities. In addition, SDG 4 (Kalimullina et al. 2021): Quality Education can be achieved by providing access to online education and digital learning resources, which can help expand access to quality education for all. SDG 5 (ElMassah and Mohieldin 2020): Gender Equality can be promoted by providing access to digital services and technologies for women and girls and addressing digital gender gaps, thereby empowering women and girls to participate fully in the digital economy and society. SDG 8 (Myovella et al. 2020): Decent Work and Economic Growth can be advanced by creating new jobs and improving the productivity of existing jobs through the use of digital technologies. SDG 9 (Nobrega et al. 2021): Industry, Innovation and Infrastructure can be advanced by driving innovation and increasing access to digital technologies in various sectors, thereby helping to spur economic growth and development. SDG 11 (Pérez-Martínez et al. 2023): Sustainable Cities and Communities can be advanced by using digital technologies to improve urban planning and management, thereby promoting more sustainable and livable communities. Finally, SDG 17 (Castro et al. 2021): Partnerships for the Goals can be advanced by fostering collaboration and sharing of knowledge and resources among various stakeholders through the use of digital technologies. Since SOSDIT has the potential to play a critical role in advancing the SDGs and promoting sustainable development, the fourth hypothesis of this study is as follows:

**H4**. *The performance of countries in SOSDIT has a positive and direct effect on their performance in achieving SDGs.*

The literature provides ample evidence that income inequality has a significant impact on achieving the SDGs. Kabeer and Santos (Kabeer and Santos 2017) argue that income inequality is often accompanied by other intersecting inequalities that can impede progress towards the SDGs. Similarly, Scherer et al. (Scherer et al. 2018) find that a reduction in income inequality is positively associated with achieving SDG 10, which aims to reduce inequalities within and among countries. Ghosh et al. (Ghosh et al. 2020) also report that reducing income inequality can contribute to achieving SDG 10 as well as SDG 11, on sustainable cities and communities, emphasizing the synergies between the goals.

In addition, Heerink and Ma (Heerink and Jia 2006) suggest that rising income inequality can lead to lower health outcomes and possibly higher fertility rates (i.e., SDG 3). Nasrollahi et al. (Nasrollahi et al. 2018) further support this by demonstrating a negative and significant relationship between income inequality and the composite index of sustainable development. Based on these findings, we hypothesize that:

**H5**. *The Gini coefficient of a country has a direct negative effect on the performance of countries in achieving the SDGs.*

**H6**. *The Gini coefficient as a moderator variable affects the process of influencing SOSDIT on the achievement of SDGs.*

The Gini coefficient, as a widely used measure of income inequality evaluation, reflects the distribution of income or consumption expenditure among individuals or households within a country. By using the Gini coefficient, we aim to measure the extent of income inequality within countries and investigate its impact on the performance of countries in achieving the SDGs. It ranges from 0 to 1, with a value of 0 indicating perfect equality (everyone has the same income) and a value of 1 indicating perfect inequality (one person has all the income). Since the larger Gini coefficient represents greater inequality, it is expected to have a negative impact on the achievement of the SDGs, which is why this issue is mentioned in the fifth hypothesis. The graphical representation of the conceptual model and research hypotheses of this study are depicted in Figure 1.

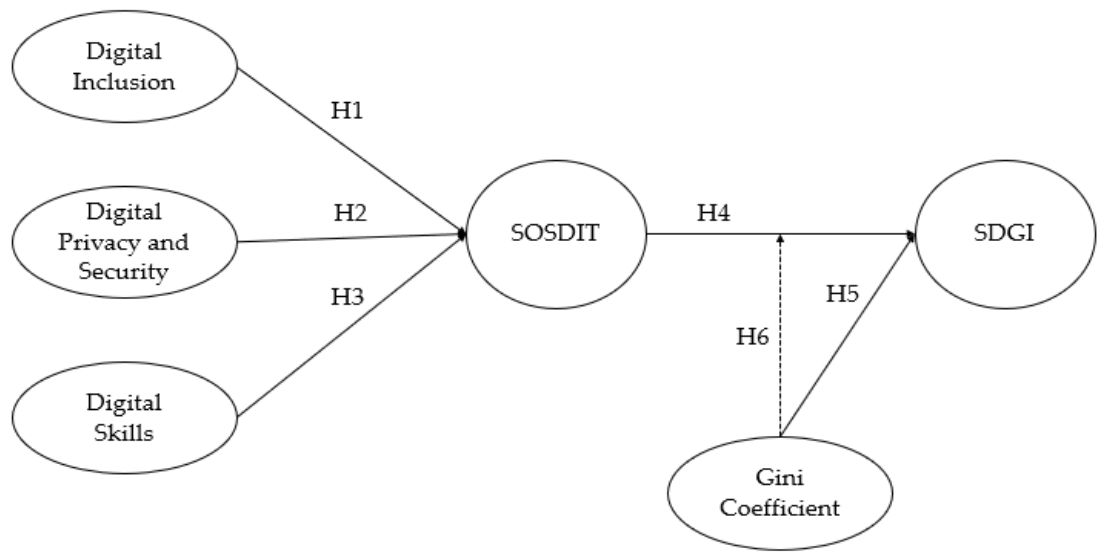

**Figure 1.** The proposed conceptual model and the hypotheses of the study.

## 3. Methodology

### 3.1. Data Source

In order to analyze the proposed conceptual model and evaluate the level of SOSDIT of the EU-27 countries, Eurostat data was used. EU-27 countries are Austria, Belgium, Bulgaria, Croatia, Cyprus, Czech Republic, Denmark, Estonia, Finland, France, Germany, Greece, Hungary, Ireland, Italy, Latvia, Lithuania, Luxembourg, Malta, Netherlands, Poland, Portugal, Romania, Slovakia, Slovenia, Spain, Sweden. In this study, the latest data available in the Eurostat database were used: it should be mentioned that the data collection was done in January 2023. In order to evaluate the digital inclusion variable, the data related to the ICT usage variable was used in this database, and the data related to ICT trust, security and privacy, and digital skills were used in this database to determine the digital privacy and security and digital skills variables, respectively. The data related to the Gini coefficient were collected from the World Development Indicators database, and finally, the SGD Index was used to evaluate the performance of the EU-27 countries in achieving the SDGs. In Table 1, the explanation of the data related to each variable is presented.

**Table 1.** Description of the data used for each of the variables of the conceptual model.

| Variables | Explanation | Question Code |
|---|---|---|
| Digital Inclusion | Use of ICT at work and activities performed | Q1-1 |
| | Work from home, from an external site or on the move | Q1-2 |
| | Internet use by individuals | Q1-3 |
| | Individuals frequently using the internet | Q1-4 |
| Digital privacy and security | Smartphone has some security system, installed automatically or provided with the operating system (individuals who used internet in the past 3 months) | Q2-1 |
| | Individuals know that cookies can be used to trace movements of people on the internet (3 months) | Q2-2 |
| | Individuals manage access to personal data on the internet (3 months): read privacy policy statements before providing personal data | Q2-3 |
| | Smartphone has some security system, installed automatically or provided with the operating system (All individuals) | Q2-4 |
| | Smartphone has some security system, installed by somebody or subscribed to it (3 months) | Q2-5 |
| | Individuals already lost information, documents, pictures or other kind of data on their smartphone as a result of a virus or other hostile type of programs (3 months) | Q2-6 |
| Digital Skills | Individuals' level of digital skills (from 2021 onwards) | Q3-1 |
| | Individuals who have used a search engine to find information | Q3-2 |
| | Individuals who have sent an email with attached files | Q3-3 |
| | Individuals who have posted messages to chat rooms, newsgroups or an online discussion forum | Q3-4 |
| | Individuals who have used the internet to make phone calls | Q3-5 |
| | Individuals who have used peer-to-peer file sharing for exchanging movies, music, etc. | Q3-6 |
| | Employed ICT specialists—total | Q3-7 |
| | Enterprises that provided training to develop/upgrade ICT skills of their personnel by NACE Rev.2 activity | Q3-8 |
| GINI Coefficient | GINI Coefficient | GINI |
| SDGs Index | SDGs Index | SDGI |

*3.2. Data Analysis*

To test proposed conceptual model in this article, structural equation modeling (SEM) based on partial least squares, which is called SEM-PLS, has been used using SmartPLS 4 software. SEM-PLS has a much better performance in evaluating models with little data (Becker et al. 2023). Since there were only 27 countries (i.e., 27 rows of data) for analysis, SEM-PLS was used.

The SEM approach in evaluating the conceptual models includes two stages. In the first stage, the measurement model is tested, and in the second stage, the structural model will be evaluated. The measurement model refers to the relationship between the observable variables (which are the questionnaire questions) and latent variables (which refers to the main variable that those questions represent). In order to evaluate the measurement model, validity and reliability tests, as well as factor analysis, are performed. This is why the structural model deals with the causal relationships between the latent variables (or the main variables of the model), and to measure the structural model, path coefficients and determination coefficients are checked.

## 4. Results

### 4.1. Measurement Model

In the present study, exploratory factor analysis was performed first, and its results are given in Table 2. It should be noted that two criteria should be considered for factor analysis: (1) the absolute value of the loading factors should be above 0.7 and (2) these loading factors should be significant in the confidence interval of at least 95% (Becker et al. 2023). The results of the factor analysis test show that all four questions selected to evaluate the latent variable of digital inclusion are above the threshold and are significant ($p < 0.05$). This is why one of the six questions considered to evaluate digital privacy and security variables is both above 0.7 and significant (i.e., Q2-2). Since the absolute values of the loading factors related to question Q2-1 and Q2-5 are significant and equal to 0.636 and 0.638, respectively (very close to the threshold value), these questions were also used in the evaluation of the final model. In other words, in the present study, the loading factor threshold was considered equal to 0.6. The negativity of the loading factor represents the inverse relationship between the question and the variable. Because this question measures the level of security that users consider when using digital tools, therefore, the more security strategies a user uses, the less security he/she feels, which is why it has an inverse relationship with the main variable. Since the loading factor of this question is significant ($p < 0.05$), we tried not to ignore the importance of this question in the proposed model. Therefore, questions Q2-1, Q2-2, Q2-5 were used to evaluate the loading variable of digital privacy and security. Finally, the loading factors of six of the eight questions assigned to evaluate digital skills are above the threshold level (which is 0.6) and are significant ($p < 0.05$). The loading factors of SDG Index and Gini coefficient variables are 1 because only one question is assigned to each of them.

**Table 2.** The result of measurement model test.

| Variables | Question Codes | Loading Factor | Sample Mean | Standard Deviation (STDEV) | T Statistics | $p$ Values |
|---|---|---|---|---|---|---|
| Digital Inclusion | Q1-1 | 0.882 | 0.88 | 0.041 | 21.369 | 0.00 |
|  | Q1-2 | 0.905 | 0.907 | 0.028 | 31.862 | 0.00 |
|  | Q1-3 | 0.941 | 0.942 | 0.015 | 63.208 | 0.00 |
|  | Q1-4 | 0.94 | 0.94 | 0.017 | 56.546 | 0.00 |
| Digital Privacy and Security | Q2-1 | −0.636 | −0.599 | 0.254 | 2.508 | 0.012 |
|  | Q2-2 | 0.750 | 0.717 | 0.171 | 4.386 | 0.00 |
|  | Q2-3 | 0.179 | 0.155 | 0.342 | 0.522 | 0.601 |
|  | Q2-4 | −0.407 | −0.365 | 0.311 | 1.31 | 0.19 |
|  | Q2-5 | 0.638 | 0.602 | 0.216 | 2.95 | 0.003 |
|  | Q2-6 | −0.446 | −0.452 | 0.18 | 2.485 | 0.013 |
| Digital Skills | Q3-1 | 0.836 | 0.829 | 0.07 | 11.864 | 0.00 |
|  | Q3-2 | 0.958 | 0.955 | 0.019 | 50.504 | 0.00 |
|  | Q3-3 | 0.914 | 0.914 | 0.023 | 39.239 | 0.00 |
|  | Q3-4 | 0.574 | 0.525 | 0.206 | 2.789 | 0.005 |
|  | Q3-5 | 0.638 | 0.601 | 0.176 | 3.631 | 0.00 |
|  | Q3-6 | 0.226 | 0.179 | 0.246 | 0.919 | 0.358 |
|  | Q3-7 | 0.895 | 0.898 | 0.03 | 29.422 | 0.00 |
|  | Q3-8 | 0.805 | 0.805 | 0.073 | 10.99 | 0.00 |
| SDG Index | SDGI | 1 | 1 | 0 | 0 | 0.00 |
| Gini Coefficient | GINI | 1 | 1 | 0 | 0 | 0.00 |

After the factor analysis, the validity and reliability of the variables were measured. The results of the reliability test show that both Cronbach's alpha and composite relia-bility (CR) of digital inclusion and digital skills are above the threshold level of 0.7. The acceptable threshold level for the average variance extracted (AVE) is 0.5. The AVE threshold value ensures that the questions assigned to a variable explain at least 50% of the variance of that variable (no other variables). Table 3 shows that the AVE values for digital inclusion and digital skills are above the threshold level of 0.5. However, the current study fails to provide the necessary reliability and validity to measure the latent variable of digital privacy and security, and this variable was removed from model—in other words, the hypothesis corresponding to this variable is rejected, which will be discussed in detail in the next part.

**Table 3.** The results of Cronbach's alpha, CR, and AVE.

| Variables | Cronbach's Alpha | CR | AVE |
|---|---|---|---|
| Digital Inclusion | 0.937 | 0.938 | 0.841 |
| Digital Skills | 0.886 | 0.94 | 0.586 |
| Digital Privacy and Security | 0.513 | 0.601 | 0.295 |

*4.2. Hypothesis Testing*

In SEM, the evaluation of the relationships between the main research variables (which are the latent variables) is called the structural model test. In the structural model test, the path coefficients should be statistically significant. The test of the structural model is actually the test of the hypotheses of this research.

Testing the First, Second, and Third Hypotheses

The first three hypotheses of this study indicate that digital inclusion, digital privacy and security, and digital skills shape the SOSDIT of a country. Of course, since the digital privacy and security variable could not achieve the required reliability and validity, despite the fact that its path coefficient ($\beta = 0.201$) is significant ($p < 0.05$) with a confidence interval of at least 95%, the second hypothesis of this study is not confirmed. Since this study employed the secondary data collected by Eurostat, failure to confirm the validity and reliability of the questions of this variable resulted in us removing the variable from the model because the authors of this article were not able to design different questions and recollect data in order to increase reliability and validity of this variable. Besides, the results of the structural model test show that the path coefficients of digital inclusion ($\beta = 0.347$) and digital skills ($\beta = 0.500$) are significant ($p < 0.05$). These results provide quantitative empirical evidence in support of the first and third hypotheses of this study. The summary of the test of the hypotheses of this research is given in Table 4.

**Table 4.** Hypothesis testing results.

| Hypotheses | β | Standard Deviation | T Statistics | $p$ Values | Result |
|---|---|---|---|---|---|
| Digital Inclusion -> SOSDIT | 0.347 | 0.024 | 14.167 | 0.00 | Confirmed |
| Digital Privacy and security -> SOSDIT | 0.201 | 0.051 | 3.929 | 0.00 | Not Confirmed |
| Digital Skills -> SOSDIT | 0.5 | 0.04 | 12.402 | 0.00 | Confirmed |
| SOSDIT -> SDG Index | 0.64 | 0.123 | 5.183 | 0.00 | Confirmed |
| Gini Coefficient -> SDG Index | −0.308 | 0.137 | 2.252 | 0.024 | Confirmed |
| Gini Coefficient x SOSDIT -> SDG Index | −0.088 | 0.141 | 0.625 | 0.532 | Not Confirmed |

The fourth hypothesis of this study refers to the effect of SOSDIT on SDG Index. The result of this hypothesis test shows that the path coefficient of this variable to the SDG Index variable ($\beta = 0.64$) is significant ($p < 0.001$). Therefore, the fourth hypothesis of this research is also confirmed. On the other hand, the fifth hypothesis of this research refers to the influence of the Gini coefficient in the process of SOSDIT affecting the SDG Index, and the present study fails to provide quantitative empirical evidence to confirm this hypothesis and this hypothesis is not confirmed. However, the results show that the Gini coefficient directly has a significant effect on the SDG Index, and since the path coefficient of this influencing process is negative (β = −0.308), the effect of this variable on the SDG Index is negative. In other words, the higher the Gini coefficient of a country, the lower its SDG Index. In Figure 2, the output of the SmartPLS software is presented, where loading factors (relationships between observable variables (yellow rectangles) and hidden variables (blue circles)), path coefficients (relationships between hidden variables), and also the magnitude of the coefficients of determination ($R^2$) (which are the same numbers written in the blue circles/hidden variables) are shown.

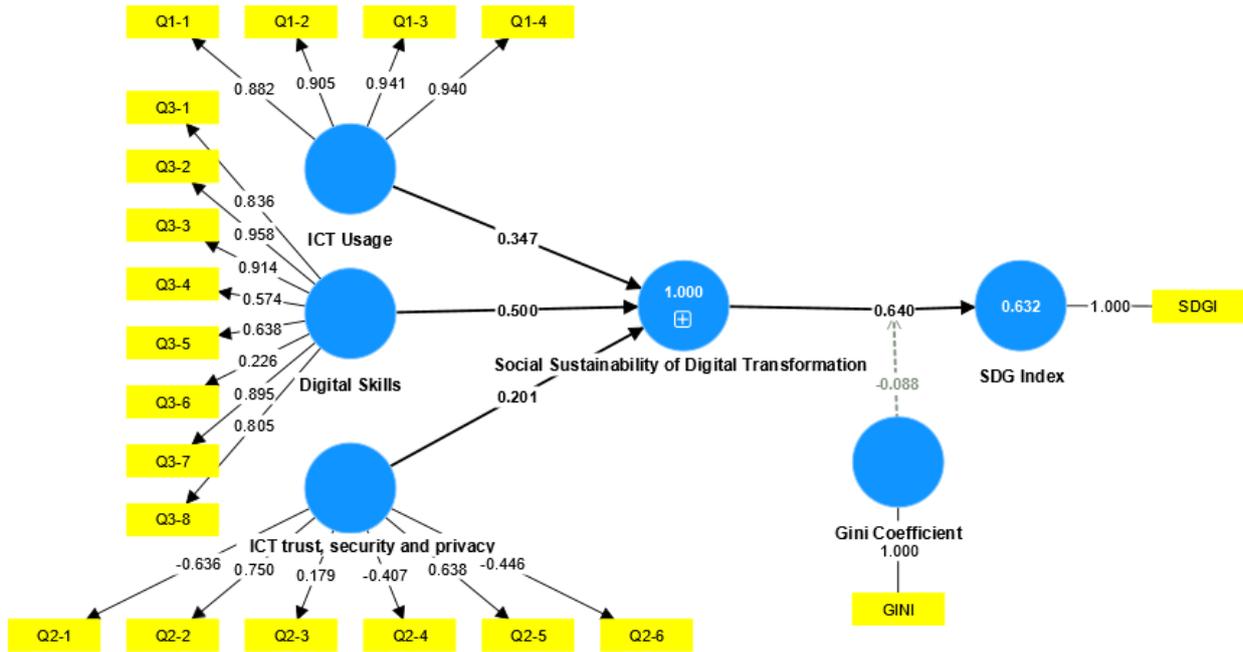

**Figure 2.** The output of SmartPLS software – the research conceptual model test.

*4.3. Answers to Research Questions*

RQ1: How socially sustainable is digital transformation in EU-27 countries?

After confirming the first and third hypotheses of this study, it is possible to calculate the SOSDIT level of EU-27 countries. The average score of countries in the field of digital inclusion and digital skills is considered as the performance of those countries in SOSDIT. The performance of the 27 European Union member states is given in Table 5 and illustrated in Figure 3. The SOSDIT score can be between 0 and 1, where 1 is the highest score that a country can achieve in terms of SOSDIT, and the number 0 represents the weakest performance of a country in SOSDIT.

**Table 5.** SOSDIT score of EU-27 countries.

| Country | SOSDIT Score | Country | SOSDIT Score |
|---|---|---|---|
| Finland | 0.59 | Sweden | 0.50 |
| Netherlands | 0.57 | Czech Republic | 0.49 |
| Denmark | 0.56 | Ireland | 0.49 |
| Austria | 0.53 | Lithuania | 0.49 |
| Germany | 0.53 | Belgium | 0.48 |
| Cyprus | 0.52 | Italy | 0.48 |
| France | 0.51 | Slovenia | 0.48 |
| Hungary | 0.51 | Poland | 0.47 |
| Luxembourg | 0.51 | Slovakia | 0.47 |
| Croatia | 0.50 | Greece | 0.46 |
| Estonia | 0.50 | Portugal | 0.45 |
| Latvia | 0.50 | Bulgaria | 0.42 |
| Malta | 0.50 | Romania | 0.41 |
| Spain | 0.50 | EU Average | 0.49 |

According to the results, Finland, Netherlands, and Denmark have obtained the highest scores in SOSDIT, of 0.59, 0.57, and 0.56, respectively, and Romania, Bulgaria, and Portugal with scores of 0.41, 0.42, and 0.45 respectively, have had the weakest performance in SOSDIT.

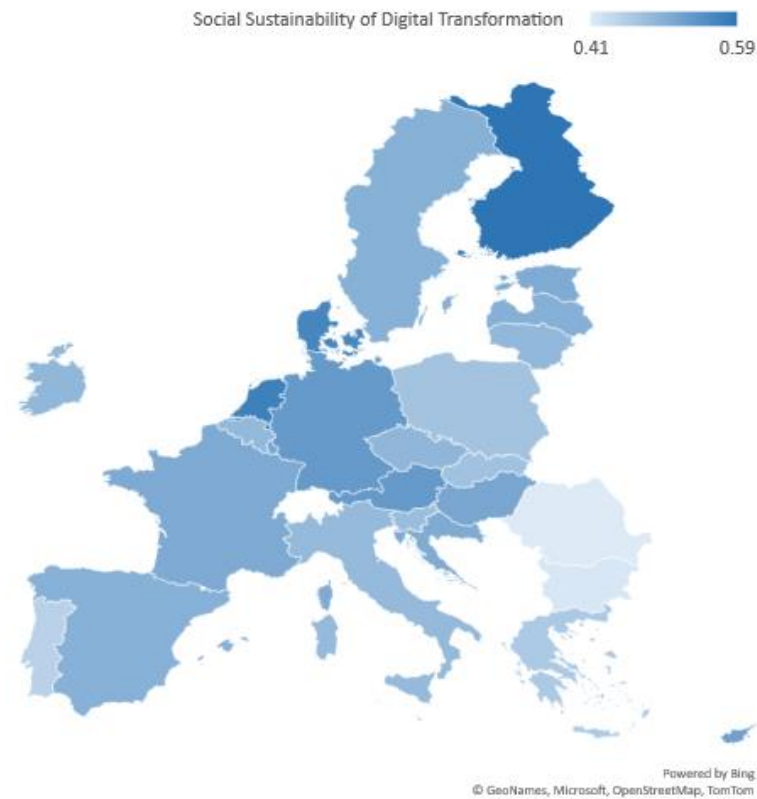

**Figure 3.** Social sustainability of digital transformation in EU-27 countries.

RQ2: Does SOSDIT lead EU-27 countries in achieving sustainable development goals?

The hypotheses of this research have been tested and hypotheses one, three, and four of this research have been confirmed, and the effect of the Gini coefficient on the SDG Index has also been proven. The magnitude and intensity of the impact of digital inclusion, digital skills, and Gini coefficient on the SDG Index variable is measured with the $R^2$ coefficient. $R^2$ = 0.632 and it illustrates that the mentioned variables can explain 63% of the changes of the SDG Index, which is a very considerable amount. On the other hand, it is suggested to check the magnitude of the F-square statistic. $F^2$ is the change in $R^2$ caused by the removal of an exogenous variable from the model. According to **Cohen** (1988), values higher than 0.15 are desirable for this statistic. The summary of $F^2$ values is given in Table 6. Table 6 shows that the $F^2$ value of the effect of Gini coefficient on SOSDIT is less than the threshold, which represents the not considerable effect of the Gini coefficient on SOSDIT.

Table 6. Results of F-square test.

| | F-Square | SDG Index |
|---|---|---|
| SOSDIT | | 0.866 |
| Gini Coefficient | | 0.174 |
| Gini Coefficient × SOSDIT | | 0.02 |

## 5. Findings and Discussion

Social sustainability of digital transformation refers to the ways in which digital technology is designed and used to support and promote social equity, fairness, and well-being, as well as to address social challenges such as inequality, poverty, and social exclusion. The development and deployment of digital skills are critical components of social sustainability as they enable individuals, organizations, and communities to participate in the digital economy and benefit from the opportunities it provides.

The confirmation of the first and third hypotheses of this study made it possible to evaluate the level of social sustainability of digital transformation in European countries. The first hypothesis of the research illustrated that digital inclusion is crucial for the social sustainability of digital transformation in a country as it guarantees equal access to the benefits and opportunities provided by technology. By ensuring digital inclusion, the benefits of digital transformation can be shared equitably among all members of society. Digital technologies have the potential to bridge existing social and economic divides, and digital inclusion helps to prevent these divides from deepening by providing equal access to technology and digital skills. Additionally, digital technologies can improve health and well-being through telemedicine and access to health information, and digital inclusion makes sure that everyone can take advantage of these benefits, regardless of their location or financial situation. Similarly, digital technologies have the power to transform education, and digital inclusion helps to guarantee that everyone has access to these benefits, regardless of their background or circumstances. Hence, digital inclusion is a critical aspect of ensuring the social sustainability of digital transformation in a country.

This study also shows that digital skills play a crucial role in the sustainability of digital transformation in a country, as they are essential for individuals, organizations, and communities to participate in the digital economy and benefit from the opportunities it provides. Without digital skills, individuals, organizations, and communities may be left behind, leading to digital inequality and exclusion. Digital skills are essential for individuals to participate in the digital economy, as many jobs now require a basic level of digital proficiency. The development of digital skills contributes to workforce development, improved access to information and services, increased entrepreneurship, and helps to bridge the digital divide by reducing digital inequality and increasing the participation of underprivileged communities and marginalized groups in the digital economy, ultimately contributing to overall economic stability and sustainability.

In addition, the present study provides evidence that the degree of SOSDIT of a country affects its performance in achieving SDGs. SOSDIT is critical to ensuring that the benefits of digital technologies are shared equitably and that the negative impacts are mitigated for all members of society, ultimately contributing to the achievement of the SDGs. Besides, it is also found that the Gini coefficient has a negative impact on a country's performance in achieving the Sustainable Development Goals (SDGs). A high Gini coefficient indicates a large divide between the wealthy and the poor, where a small percentage of the population controls a large proportion of the wealth. This leads to several negative outcomes for the country's SDG Index. Firstly, it creates poverty and hardship for a large portion of the population, negatively impacting the SDGs of No Poverty and Reduced Inequalities. Secondly, it can result in decreased economic growth as the purchasing power of the majority of the population is reduced, negatively affecting the SDG of Decent Work and Economic Growth. Thirdly, it leads to inadequate access to basic services such as healthcare, education, clean water, and sanitation. Finally, high levels of income inequality can cause political instability, which can negatively impact a country's ability to achieve the SDGs. Thus, it is important for countries to address income inequality through policies that promote equitable distribution of wealth and resources, contributing to a more sustainable future.

To encapsulate, the theoretical contributions of this study lie in the development of a conceptual model to evaluate the SOSDIT among EU-27 countries and the examination of the relationship between SOSDIT, the performance of countries in achieving the SDGs, and the Gini coefficient. The study provides a framework for understanding the importance of digital inclusion and digital skills as the building blocks of SOSDIT and highlights the direct impact of SOSDIT on the performance of countries in achieving the SDGs. The study also sheds light on the negative effect of the Gini coefficient on the performance of countries in achieving the SDGs. These contributions add to the existing literature on digital transformation, social sustainability, and the SDGs and provide valuable insights for policymakers, researchers, and practitioners.

Based on the findings of this study, it is clear that an interdisciplinary approach is necessary to understand the complex relationship between digital inclusion, sustainability, and development. This study draws on insights from multiple fields, including economics, information systems, and sustainability studies. The developed conceptual model integrates these perspectives and provides a framework for analyzing the relationship between digital inclusion and the achievement of the SDGs.

One critical reflection that this study systematizes is the need to view digital inclusion as a fundamental component of social sustainability. This view challenges traditional notions of sustainability that focus exclusively on environmental sustainability and recognizes that sustainable development must also include social and economic sustainability. This study shows that digital inclusion is a key enabler of social sustainability and can play a crucial role in achieving the SDGs.

Another critical reflection that our study systematizes is the importance of recognizing the role of inequality in shaping the relationship between digital inclusion and sustainable development. These findings show that the Gini coefficient has a significant negative effect on the performance of countries in achieving the SDGs, highlighting the need to address inequality as part of efforts to promote sustainable development. This study underscores the importance of taking an intersectional approach that recognizes the ways in which different forms of inequality intersect and compound one another.

*5.1. Theoretical Contributions*

The theoretical contributions of this manuscript are multi-fold. Firstly, the concept of SOSDIT is introduced, which focuses on the ways in which digital technology can be designed and used to promote social equity, fairness, and well-being, and to address social challenges such as inequality, poverty, and social exclusion. This concept highlights the need to prioritize social sustainability in the design and deployment of digital technologies, which can help to ensure that the benefits of digital transformation are shared equitably among all members of society and that the negative impacts are mitigated for all.

Secondly, the study identifies digital inclusion and digital skills development as critical components of SOSDIT. Digital inclusion refers to the need to ensure equal access to the benefits and opportunities provided by technology, while digital skills development is essential for individuals, organizations, and communities to participate in the digital economy and benefit from the opportunities it provides. These concepts highlight the importance of ensuring that all members of society have access to digital technologies and the skills needed to use them effectively, which can help to prevent the deepening of social and economic divides and contribute to the achievement of the SDGs.

Thirdly, the study provides empirical evidence of the relationship between SOSDIT and the achievement of SDGs. The findings demonstrate that countries with a higher degree of SOSDIT have a higher performance in achieving SDGs, indicating the im-portance of prioritizing social sustainability in the design and deployment of digital technologies. This provides a theoretical basis for policymakers to develop policies and strategies to promote social sustainability in the digital transformation process, ultimately contributing to a more sustainable future.

Finally, the study identifies the negative impact of income inequality, as measured by the Gini coefficient, on a country's performance in achieving SDGs. This highlights the need to address income inequality through policies that promote equitable distribution of wealth and resources, which can contribute to achieving SDGs and promoting a more sustainable future.

*5.2. Practical Contributions*

The practical contributions of this study are significant as it provides important insights for policymakers, researchers, and practitioners to promote social sustainability in the digital transformation process.

Firstly, the study highlights the importance of digital inclusion and digital skills as critical components of social sustainability. Policymakers can use these findings to design and implement policies that ensure equal access to technology and digital skills for all members of society, regardless of their background or circumstances. This can be achieved through initiatives such as free digital skills training programs, subsidized access to technology, and policies that ensure the availability of digital services in rural and underprivileged areas.

Secondly, the study emphasizes the need to address income inequality through policies that promote equitable distribution of wealth and resources. Policymakers can use these findings to design policies that address income inequality, such as progressive taxation, social welfare programs, and investment in education and training.

Thirdly, the study highlights the direct impact of SOSDIT on the performance of countries in achieving the SDGs. Policymakers can use these findings to prioritize SOSDIT in their national development plans, allocate resources to promote digital inclusion and digital skills, and design policies that ensure that the benefits of digital transformation are shared equitably among all members of society.

## 6. Conclusions

In conclusion, this study highlights the critical role of SOSDIT in the achievement of the Sustainable Development Goals (SDGs) among EU-27 countries. Our findings demonstrate that digital inclusion and digital skills are the main factors of SOSDIT and significantly impact the ability of countries to attain SDGs. Our study also highlights the negative impact of income inequality, as measured by the Gini coefficient, on the performance of countries in achieving SDGs. These findings have important implications for policymakers and decision-makers, as they suggest that investment in digital inclusion and digital skills can have a positive impact on the sustainability of digital transformation and contribute to the achievement of the SDGs. Furthermore, reducing income inequality through progressive taxation, investment in education and job training programs, inclusive economic growth, and safety net programs for the most vulnerable populations, can also contribute to a more sustainable digital transformation and improved performance in achieving the SDGs. This study underscores the need for continued research and action to ensure a socially sustainable digital transformation that benefits all individuals and contributes to a more sustainable future.

This study demonstrates the value of an interdisciplinary approach to understanding the complex relationship between digital inclusion, sustainability, and development. By systematically integrating insights from multiple fields, this study provides a framework for analyzing the relationship between digital inclusion and the achievement of the SDGs, highlighting the need to view digital inclusion as a fundamental component of social sustainability, and emphasizing the importance of addressing inequality in efforts to promote sustainable development.

*6.1. Practical Implications and Recommendations*

- Invest in digital infrastructure: Governments should invest in the development of digital infrastructure, such as high-speed internet access, to ensure that everyone has access to technology and digital skills.
- Provide digital skills training: Governments should provide training and support to ensure that everyone has the skills and knowledge to use technology effectively. This includes training for individuals and organizations, as well as training for educators to ensure that digital skills are taught in schools.
- Promote digital literacy: Governments should promote digital literacy and ensure that individuals have the skills and knowledge to use technology effectively. This can be achieved through education and training programs, as well as through public awareness campaigns.
- Foster digital inclusion: Governments should foster digital inclusion by addressing issues such as the digital divide and ensuring that everyone has access to technology and digital skills. This can be achieved through public–private partnerships and community initiatives.
- Investment in education and job training programs: Providing access to education and job training programs can help to equip individuals with the skills needed to secure well-paying jobs, increase their earning potential, and reduce income inequality. This can also lead to a reduction in the Gini coefficient and improve a country's performance in achieving the SDGs, especially the SDGs of Decent Work and Economic Growth, No Poverty, and Quality Education. By investing in education and job training programs, a country can provide opportunities for individuals to improve their lives and contribute to a more sustainable future.

*6.2. Recommendations for Future Research*

As with any study, there are limitations to the scope of research and the available data. This study provides valuable insights into the relationship between SOSDIT, the performance of countries in achieving the SDGs, and the Gini coefficient among EU-27 countries. However, there is a need for further research to build on these findings and to expand the understanding of this relationship beyond the EU-27. In this section, recommendations for future research are outlined that could help to address these limitations and provide a more comprehensive understanding of the role of digital transformation in achieving social sustainability and the SDGs. These recommendations include investigating regional disparities, examining the impact of technology adoption, studying the role of the private sector, and evaluating the effectiveness of policy interventions.

- Investigation of regional disparities: This study focuses on EU-27 countries, but future research could explore regional disparities within countries and how they affect the performance of countries in achieving the SDGs.
- Examination of the impact of technology adoption: Future research could explore the impact of technology adoption, such as the adoption of artificial intelligence and the internet of things, on SOSDIT and the performance of countries in achieving the SDGs.
- Study of the role of the private sector: The private sector plays a critical role in digital transformation and the achievement of the SDGs. Future research could explore the role of the private sector in promoting SOSDIT and contributing to the achievement of the SDGs.
- Evaluation of the effectiveness of policy interventions: Future research could evaluate the effectiveness of policy interventions aimed at promoting SOSDIT and the performance of countries in achieving the SDGs. This would provide valuable insights into what works and what does not, helping policymakers and decision-makers to make informed decisions.

*6.3. Limitations of the Study*

While this study provides valuable insights into the relationship between SOSDIT and the performance of countries in achieving the SDGs, it is important to note several limitations that may affect the interpretation of the results.

Firstly, the study only focuses on EU-27 countries, which may limit the generalizability of the findings to other regions or countries. Future research could examine the applicability of these findings to other regions and expand the scope of analysis beyond the EU-27.

Secondly, the study relies on available indicators to measure SOSDIT, the performance of countries in achieving the SDGs, and the Gini coefficient. The limitations of these indicators should be considered when interpreting the results, and future studies could explore alternative or additional indicators to provide a more comprehensive assessment of these concepts.

Lastly, the study is limited by the availability and quality of data, as well as potential measurement errors or biases. Further research could address these limitations by collecting more comprehensive and accurate data, using alternative measurement approaches, or conducting case studies to provide a more nuanced understanding of the relationship between SOSDIT and the achievement of the SDGs.

**Author Contributions:.** Conceptualization, S.N. and Sz.H.; methodology, T.A.; software, S.N.; validation, Sz.H.; formal analysis, S.N., T.A.; investigation, S.N.; data curation, T.A.; writing—original draft preparation, S.N., T.A.; writing—review and editing, Sz.H.; visualization, S.N.; supervision, Sz.H.

**Funding:** This research received no external funding.

**Institutional Review Board Statement:** Not applicable.

**Informed Consent Statement:** Not applicable.

**Data Availability Statement:** Data was obtained from Eurostat and World Bank datasets that are available in the following links: https://ec.europa.eu/eurostat/databrowser/explore/all/science?lang=en&subtheme=isoc&display=list&sort=category,https://eu-dashboards.sdgindex.org/profiles,, and https://data.worldbank.org/indicator/SI.POV.GINI?most_recent_value_desc=false

**Conflicts of Interest:** The authors declare no conflict of interest.